\documentclass[aps,prd,preprint,superscriptaddress,nofootinbib,floatfix]{revtex4-1} 

\usepackage{epsfig}
\usepackage{graphicx}
\usepackage[normalem]{ulem}
\usepackage{amssymb}
\usepackage{amsmath}
\usepackage{amsfonts}
\usepackage{latexsym}
\usepackage{verbatim}
\usepackage{mathtools}
\usepackage{setspace}
\usepackage{slashed}
\usepackage[all]{xypic}
\usepackage{psfrag}
\usepackage{comment}
\usepackage{array}
\usepackage{amssymb}
\usepackage{amsmath}
\usepackage{amsthm}
\usepackage{float}
\usepackage{amsfonts,wasysym,epsfig,verbatim,subfigure,bm,mathrsfs,lipsum}
\usepackage[colorlinks]{hyperref}
\usepackage{textcomp}

\begin{document}
	
\title{ Parity violation in stochastic gravitational wave background from inflation }

\author{ Rong-Gen Cai }
\email{cairg@itp.ac.cn}
\affiliation{ CAS Key Laboratory of Theoretical Physics, Institute of Theoretical Physics, Chinese Academy of Sciences, Beijing 100190, China }
\affiliation{ School of Physical Sciences, University of Chinese Academy of Sciences, Beijing 100049, China } 

\author{ Chengjie Fu }
\email[Corresponding author:~]{fucj@itp.ac.cn}
\affiliation{ CAS Key Laboratory of Theoretical Physics, Institute of Theoretical Physics, Chinese Academy of Sciences, Beijing 100190, China }

\author{ Wang-Wei Yu }
\email[Corresponding author:~]{yuwangwei@mail.itp.ac.cn}
\affiliation{ CAS Key Laboratory of Theoretical Physics, Institute of Theoretical Physics, Chinese Academy of Sciences, Beijing 100190, China }
\affiliation{ School of Physical Sciences, University of Chinese Academy of Sciences, Beijing 100049, China } 
	
\begin{abstract}
We study the inflationary implications of a novel parity-violating gravity model, which modifies the teleparallel equivalent of general relativity by introducing the Nieh-Yan term coupled to an axion-like field. The parity-violating Nieh-Yan term results in the velocity birefringence of gravitational waves (GWs) and triggers the tachyonic instability only for one of the two circular polarization states. We consider that the inflaton is identified as
the coupled axion-like field with a wiggly potential characterized by steep cliffs connected by smooth plateaus. During inflation, the temporary fast roll of axion on the cliff-like region leads to the significant enhancement of the tensor perturbations in one polarization state with the wave numbers that exit the horizon around this period. In this setup, the resulting energy spectrum for GWs presents a sizable localized bump involving the contribution of only one polarization state. This chiral GW background is detectable by LISA and Taiji, and its chirality can be determined by correlating two detectors, which provide an opportunity to probe the inflation and test the gravity model.
\end{abstract}

\maketitle

\section{Introduction}
The cosmic inflation, an early period of accelerated expansion of the Universe, is an appealing scenario not only for resolving the horizon and flatness problems of standard big-bang cosmology, but also for explaining the origin of the primordial density perturbations that are the seeds for the temperature anisotropies of the cosmic microwave background (CMB) \cite{Starobinsky:1980te,Guth:1980zm,Linde:1981mu,Mukhanov:1981xt,Guth:1982ec,Starobinsky:1982ee}. Thanks to the precise measurements of CMB, we have gained much information about the primordial fluctuations at large scales with comoving wave numbers $k \lesssim 1{\rm Mpc}^{-1}$, e.g. the amplitude of power spectrum for the scalar perturbations $\mathcal{P}_\zeta(k_{\rm cmb}) \simeq 2.1 \times 10^{-9}$ \cite{Planck:2018jri} and the tensor-to-scalar ratio $r\equiv \mathcal{P}_h(k_{\rm cmb})/\mathcal{P}_\zeta(k_{\rm cmb}) < 0.06$ \cite{BICEP2:2018kqh} with a pivot scale $k_{\rm cmb}=0.05{\rm Mpc}^{-1}$. In contrast, very little is known about the primordial fluctuations on small scales ($k \gtrsim 1{\rm Mpc}^{-1}$) due to the limited sensitivity. Therefore, for now one can only probe a small fraction of inflation through CMB. But fortunately, the remarkable success in the direct detection of gravitational waves (GWs) by the LIGO/Virgo collaboration \cite{LIGOScientific:2018mvr,LIGOScientific:2020ibl} opens up a new doorway to detect the early Universe. Operational and upcoming GW experiments (e.g. aLIGO \cite{LIGOScientific:2014pky}, LISA \cite{LISA:2017pwj}, Taiji \cite{Ruan:2018tsw}) hold the potential to complete the picture of inflation in the near future.

From the observational side, there have been many works attempting to explore the inflationary models that predict an observable stochastic GW background on small scales through the various mechanisms, such as small scale enhancement of tensor perturbations \cite{Biagetti:2013kwa,Fujita:2014oba,Cai:2016ldn,Mylova:2018yap,Satoh:2007gn,Fu:2020tlw}, emission of GWs by particle production during inflation \cite{Cook:2011hg,Goolsby-Cole:2017hod,Barnaby:2011qe,Garcia-Bellido:2016dkw,Ozsoy:2020ccy,Almeida:2020kaq,Ozsoy:2020kat}, and enhanced curvature perturbations at small scales as a source of GWs during radiation-dominated era \cite{Saito:2008jc,Bugaev:2010bb,Inomata:2016rbd,Di:2017ndc,Bartolo:2018evs,Cai:2019jah,Fu:2019vqc,Cai:2019bmk,Fu:2020lob,Yi:2020cut,Cai:2021wzd,Gao:2020tsa,Kawai:2021edk}, see also \cite{Bartolo:2016ami,Caprini:2018mtu,Bian:2021ini} for reviews. If such a GW signal is indeed detected, the location, shape and amplitude of the predicted GW spectrum can provide experimental evidence on these inflationary scenarios. In addition, the parity violation in the stochastic GW background is a very distinctive property for unambiguously differentiating the models producing a chiral GW signal from others. 

The chiral GW signals can be generated either from the parity-violating sources or in the parity-violating gravity theories, where two circular polarization states of GWs propagate with different behaviors. The former case has been extensively investigated in the axion inflation with a coupling of the axion to gauge field through the Chern-Simons term \cite{Barnaby:2011qe,Garcia-Bellido:2016dkw,Ozsoy:2020ccy,Almeida:2020kaq,Ozsoy:2020kat}, in which the motion of the axion causes the amplification of one helicity state of gauge field. We will consider the latter scenario in this study. A well-studied parity-violating gravity theory is the Chern-Simons gravity, and its inflationary applications have been explored extensively in the literature such as \cite{Satoh:2007gn,Fu:2020tlw,Alexander:2004us,Kawai:2017kqt,Odintsov:2021kup}, in which Refs. \cite{Satoh:2007gn,Fu:2020tlw,Odintsov:2021kup} predict the observable chiral GW backgrounds. 
Despite the fact that the Chern-Simons gravity is the simplest parity-violating gravity in the metric theory based on Riemannian geometry, it suffers from the ghost-instability problem owing to the presence of higher derivative terms. On the basis of Chern-Simons gravity, the more complex parity-violating theories without Ostrogradsky ghost have been explored in Ref. \cite{Crisostomi:2017ugk} by including first or higher derivatives of the coupling scalar field.

Recently, Li \textit{et al.} \cite{Li:2020xjt,Li:2021wij} have developed a simple and healthy parity-violating gravity model based on the teleparallel gravity. In this model, the teleparallel equivalent of general relativity (TEGR), which is dynamically equivalent to the general relativity (GR),
is modified by incorporating the scalar field coupled Nieh-Yan term into its action, which violates the parity symmetry in gravity. In contrast to the Chern-Simons gravity, such a Nieh-Yan modified TEGR does not introduce any higher derivative terms and is free from the ghost instability. By using the observational data from LIGO-Virgo GW events, the first constraint on this model has been performed in \cite{Wu:2021ndf}. In this paper, we focus on the phenomenological implications of this model during inflation. Given that the measurement of chirality for an isotropic GW background will reach enough high sensitivity by correlating LISA and Taiji \cite{Seto:2020zxw,Orlando:2020oko}, we expect the stochastic GW background generated from inflation in the Nieh-Yan modified TEGR to be peaked at the frequencies accessible to the LISA-Taiji network.

The organization of this paper is as follows.
In the next section, we briefly review the Nieh-Yan modified TEGR and present its field equations in the spatially flat Friedmann-Robertson-Walker (FRW) cosmology. In Sec. \ref{sec3}, we apply this model to the axion inflation and study an explicit numerical example of producing an observable chiral GW background. Section \ref{sec4} is devoted to our conclusions. In addition, the primordial scalar perturbations generated through the curvaton scenario will be presented in Appendix \ref{appendix}.

Throughout this paper, we employ natural units $c=\hbar\equiv1$ and set the reduced Planck mass $M_{\rm p}=(8\pi G)^{-1/2}$ equal to one.

\section{Nieh-Yan modified teleparallel equivalent of general relativity}
The teleparallel theories of gravity have the tetrad $e^A_{~\mu}$ and the spin connection $\omega^A_{~B\mu}$ as their fundamental variables, where we choose $A, B, ... (= 0,1,2,3)$ and $a, b, ... (= 1,2,3)$ to denote the indices of the tangent space that is by definition a Minkowski spacetime, while $\mu, \nu, ... (= 0,1,2,3)$ and $i, j, ... (= 1,2,3)$ for the spacetime indices. It should be emphasized that the spin connection describes the inertial effects present in a given frame and is not an independent variable from the tetrad. In the teleparallel gravity, the tetrad can be used to construct the spacetime metric via the relation, 
\begin{align}
g_{\mu\nu}=\eta_{AB}e^A_{~\mu} e^B_{~\nu}\,,
\end{align}
where $\eta_{A B}={\rm diag}(1,-1,-1,-1)$ denotes the Minkowski metric, and the spin connection must satisfy the vanishing curvature condition reading
\begin{align}
\hat{R}^A_{~B\mu\nu} = \partial_\mu\omega^A_{~B\nu} - \partial_\nu\omega^A_{~B\mu} + \omega^A_{~C\mu}\omega^C_{~B\nu} - \omega^A_{~C\nu}\omega^C_{~B\mu}\equiv0\,.
\end{align}
From the tetrad and the spin connection, one can define the teleparallel connection as
\begin{align}
\hat\Gamma^\sigma_{~\mu\nu} = e_A^{~\sigma}( \partial_\mu e^A_{~\nu} + \omega^A_{~B\mu} e^B_{~\nu} )\,,	
\end{align}
which leads to a vanishing Riemann tensor,
\begin{align}
\hat{R}^\sigma_{~\rho\mu\nu} = \partial_\mu \hat\Gamma^\sigma_{~\rho\nu} - \partial_\nu \hat\Gamma^\sigma_{~\rho\mu} + \hat\Gamma^\sigma_{~\alpha\mu} \hat\Gamma^\alpha_{~\rho\nu} - \hat\Gamma^\sigma_{~\alpha\nu} \hat\Gamma^\alpha_{~\rho\mu} = 0\,,
\end{align}
but a nonzero torsion tensor,
\begin{align}
T^\sigma_{~\mu\nu} =  \hat\Gamma^\sigma_{~\mu\nu} - \hat\Gamma^\sigma_{~\nu\mu}\,.
\end{align}

Under a local Lorentz transformation $\Lambda^A_{~B}(x^\mu)$, the tetrad and the spin connection change according to
\begin{align}
e^A_{~\mu} \mapsto \Lambda^A_{~B} e^B_{~\mu}\,, \quad \omega^A_{~B\mu} \mapsto \Lambda^A_{~C}(\Lambda^{-1})^D_{~B} \omega^C_{~D\mu} + \Lambda^A_{~C}\partial_\mu (\Lambda^{-1})^C_{~B}\,,
\end{align}
and it follows that the metric $g_{\mu\nu}$ and the teleparallel connection $\hat\Gamma^\sigma_{~\mu\nu}$ (as well as the Riemann tensor $\hat{R}^\sigma_{~\rho\mu\nu}$ and the torsion tensor $T^\sigma_{~\mu\nu}$) are invariant. If starting from an inertial frame with vanishing spin connection, the spin connection in different classes of noninertial frames, obtained by performing local Lorentz transformation $\Lambda^A_{~B}(x^\mu)$, takes the following simple form,
\begin{align}
\omega^A_{~B\mu} = \Lambda^A_{~C}\partial_\mu (\Lambda^{-1})^C_{~B}\,,
\end{align}
from which we will replace $\omega^A_{~B\mu}$ with $\Lambda^A_{~B}$ as the fundamental variable in the action principle.

The gravitational action in the TEGR is given by
\begin{align}\label{TEGR_action}
S_{\rm TEGR} = \frac{1}{2}\int d^4x~e~ T = \int d^4x~e \left( - \frac{1}{2} T^\rho_{~\rho\mu}T^{\sigma~\mu}_{~\sigma} + \frac{1}{8} T^{\rho\sigma\mu} T_{\rho\sigma\mu}  + \frac{1}{4} T^{\mu\sigma\rho} T_{\rho\sigma\mu} \right) \,,
\end{align}
where $e=\sqrt{-g}$ is the determinant of the tetrad and $T$ denotes the torsion scalar. It is interesting to mention that the torsion scalar $T$ is related to the Ricci scalar $R$, defined by the Levi-Civita connection associated with the metric, by
\begin{align}
T = -R - \frac{2}{e}\partial_\mu\left(e T^{\rho\mu}_{~~~\rho}\right)\,,
\end{align}
so that the action \eqref{TEGR_action} is identical to the Einstein-Hilbert action up to a surface term,
\begin{align}\label{TEGR}
S_{\rm TEGR} = \int d^4x \left[ - \frac{\sqrt{-g} }{2} R - \partial_\mu\left(e T^{\rho\mu}_{~~~\rho}\right) \right] \,.
\end{align} 
In this way, GR and TEGR produce identical field equations and are dynamically equivalent to each other.
On the basis of the TEGR action, the model proposed in \cite{Li:2020xjt} introduces the Nieh-Yan term coupled to an axion-like field, which takes the form
\begin{align}\label{NY}
S_{\rm NY} = \int d^4x~e~\frac{c\theta}{4} T_{A\mu\nu}\tilde T^{A\mu\nu} 
\end{align} 
in the framework of teleparallel gravity.
Here, the axion-like $\theta$ field is a dynamical scalar field, $c$ represents the coupling constant, $T^A_{~\mu\nu}=  e^A_{~\sigma}T^\sigma_{~\mu\nu}$ and $\tilde T^{A\mu\nu}=(1/2)\varepsilon^{\mu\nu\rho\sigma}T^A_{~\rho\sigma}$ are respectively the torsion two form and its dual with the Levi-Civita tensor $\varepsilon^{\mu\nu\rho\sigma}$ being defined in terms of the antisymmetric symbol $\epsilon^{\mu\nu\rho\sigma}$ as $\varepsilon^{\mu\nu\rho\sigma}=\epsilon^{\mu\nu\rho\sigma}/\sqrt{-g}$.  
Adding the kinetic and potential terms of the $\theta$ field, the full action for the Nieh-Yan modified TEGR is given by
\begin{align}\label{action}
S  = \int d^4 x \sqrt{-g}\left[- \frac{R}{2} + \frac{c\theta}{4} \mathcal{T}_{A\mu\nu} \tilde{\mathcal{T}}^{A \mu \nu} + \frac{1}{2}\nabla_\mu \theta\nabla^\mu \theta -V(\theta) \right] + S_m,
\end{align}
where the surface term in Eq. \eqref{TEGR} has been disregarded, and the action $S_m$ describes the other matter minimally coupled to gravity through the metric.

Varying the action with respect to $e^A_{~\mu}$ and $\Lambda^A_{~B}$ separately yields the following gravitational field equation and constrain equation, 
\begin{align}
G^{\mu\nu} + N^{\mu\nu} &= T^{\mu\nu} + T_\theta^{\mu\nu}\,, \label{Einstein_equation} \\
N^{\mu\nu} &= N^{\nu\mu}\,, \label{symmetry_N}
\end{align}
where $G^{\mu\nu}$ is the Einstein tensor, $N^{\mu\nu} = c e_A^{~\nu} \partial_\rho \theta \tilde{T}^{A\mu\rho}$, $T^{\mu\nu} = -(2/\sqrt{-g})\delta S_m / \delta g_{\mu\nu}$ and $T_\theta^{\mu\nu} = [V(\theta) - \nabla_\alpha \theta \nabla^\alpha \theta / 2]g^{\mu\nu} + \nabla^\mu \theta \nabla^\nu \theta$ are the energy-momentum tensors for the matter and the $\theta$ field, respectively. Then, we perform the variation of the action with respect to the $\theta$ field, and obtain its Klein-Gordon equation,
\begin{align}
\square \theta + V_\theta - \frac{c}{4}\mathcal{T}_{A\mu\nu} \mathcal{\tilde T}^{A\mu\nu} = 0\,,
\end{align}
where a subscript $\theta$ denotes $\partial/\partial\theta$.

In this paper, we work with the spatially flat FRW spacetime, characterized by the line element $ds^2=a(\eta)^2\left(d\eta^2 - \delta_{ij}dx^idx^j \right)$ with $\eta$ being the conformal time. Using the spherical coordinates $(\eta,r,\vartheta,\varphi)$, we can choose a nondiagonal tetrad as \cite{Krssak:2018ywd}
\begin{align}\label{tetrad}
e^A_{~\mu} = a(\eta)\left(               
\begin{array}{cccc}  
	1 & 0 & 0 & 0\\ 
	0 & \sin\vartheta\cos\varphi & r\cos\vartheta\cos\varphi & -r\sin\vartheta\sin\varphi \\  
	0 & \sin\vartheta\sin\varphi & r\cos\vartheta\sin\varphi & r\sin\vartheta\cos\varphi \\  
	0 & \cos\vartheta & -r\sin\vartheta & 0 \\  
\end{array}
\right)\,,
\end{align}
which corresponds to zero spin connection, $\omega^A_{~B\mu}=0$. It follows that the background evolution is dominated by the following equations,
\begin{align}\label{BG_Eq}
3\mathcal{H}^2 = a^2\left( \rho_\theta + \rho \right)\,, \quad - 2 \mathcal{H}' - \mathcal{H}^2 =  a^2 \left( p_\theta + p\right)\,, \quad \theta'' + 2 \mathcal{H}\theta' + a^2V_{\theta} = 0\,,
\end{align}
where $\mathcal{H}\equiv a^\prime/a = aH$ is the conformal Hubble parameter and a prime denotes the derivative with respect to the conformal time. Here, $\rho_\theta = \dot\theta^2/2 + V$ and $p_\theta = \dot\theta^2/2 - V$, in which the derivatives with respect to the cosmic time have been denoted by a dot, respectively represent the energy density and the pressure of the $\theta$ field, and $\rho$ and $p$ are  energy density and pressure associated with the matter. From these equations, it is clear that the Nieh-Yan term has no effect on the background evolution. 

Let us now turn to the linear cosmological perturbations. We may convert the tetrad \eqref{tetrad} to the Cartesian coordinates $(\eta,x,y,z)$ and get a diagonal tetrad written as $e^A_{~ \mu}={\rm diag} (a,a,a,a)$ for convenience.
So that one can parametrize the components of the perturbed tetrad by \cite{Izumi:2012qj,Golovnev:2018wbh} 
\begin{align}
e^0_{~0} & = a(1+\phi)\,, \quad e^0_{~i} = a\partial_i \beta\,, \quad e^a_{~0} = a\delta_{ai}\partial_i \gamma\,, \nonumber  \\
e^a_{~i} & = a \delta_{aj}\left[(1-\psi)\delta_{ij} + \partial_i \partial_j E + \epsilon_{ijk}\partial_k \lambda + \frac{1}{2} h_{ij} \right]\,,
\end{align}
where $\phi$, $\beta$, $\gamma$, $\psi$, $E$, and $\lambda$ are the scalar perturbations, and $h_{ij}$ denotes the transverse-traceless tensor perturbation. Note that here  we ignore the vector perturbations of disinterest in this paper. The corresponding metric components can be expressed as
\begin{align}
g_{00} & = a^2 (1+2\phi)\,,  \quad  g_{0i} = - a^2\partial_iB\,, \nonumber \\
g_{ij} & = -a^2\left[(1-2\psi)\delta_{ij} + 2 \partial_i \partial_j E  + h_{ij} \right],
\end{align}
where $B\equiv\gamma - \beta$. 
To describe the field fluctuations, we decompose the $\theta$ field into homogeneous and fluctuating parts: $\theta(\eta,\bf{x})=\theta(\eta) + \delta\theta(\eta,\bf{x})$. Throughout we adopt the convenient notation that $\delta X$ denotes the perturbation to the homogeneous quantity $X$. Considering that the matter is a perfect fluid, its energy-momentum tensor up to linear order can be expressed as
\begin{align}
T^0_0  = \rho + \delta \rho\,, \quad T^0_i = - (\rho + p)\partial_i v\,, \quad T^i_j = - (p + \delta p) \delta_{ij}\,,
\end{align}
where $v$ is a scalar velocity potential. Firstly, the symmetry of $N^{\mu\nu}$ described by Eq. \eqref{symmetry_N} yields the following constraint on the scalar sector,
\begin{align}
\theta^\prime \psi + \mathcal{H}\delta\theta = 0\,.
\end{align}
This implies that the gauge-invariant variable $\zeta_\theta = -( \psi + \mathcal{H}\delta\theta/\theta^\prime)$, which is used to describe the scalar perturbations, is zero identically.
Then, taking the spatially flat gauge, $\psi=E=0$ (and hence $\delta\theta=0$), the first-order perturbed Einstein equations derived from Eq. \eqref{Einstein_equation} are given by
\begin{align}
6\mathcal{H}^2\phi + 2\mathcal{H}\nabla^2B &=   \theta^{\prime2} \phi  - a^2 \delta\rho\,,  \nonumber \\ 
2\mathcal{H} \phi &= - a^2(\rho + p)v\,,  \nonumber \\ 
2\mathcal{H} \phi^\prime + (4\mathcal{H}^\prime + 2\mathcal{H}^2) \phi  &= - \theta^{\prime2} \phi + a^2\delta p \,.  
\end{align}
On the other hand, the tensor perturbation, $h_{ij}$, obeys the following equation of motion, 
\begin{align}\label{EoM_h}
h_{ij}^{\prime\prime} + 2\mathcal{H}h_{ij}^\prime - \nabla^2 h_{ij} + \frac{1}{2}c\theta^\prime\left( \epsilon_{lki}\partial_l h_{jk} + \epsilon_{lkj}\partial_l h_{ik} \right) = 0\,,
\end{align}
where the last term arising from the action \eqref{NY} results in the parity violation in the tensor perturbations.
To analyze this phenomenon, it is convenient to decompose the tensor perturbations in terms of the circular polarization bases,
\begin{align}\label{Fourier}
h_{ij}(\eta,{\bf x}) = \sum_{A=R,L} \int \frac{d{\bf k}^3}{(2\pi)^{3/2}} h^A(\eta,{\bf k}) e_{ij}^A({\bf k}) e^{i {\bf k} \cdot {\bf x}}\,,
\end{align}
where $R$ and $L$ respectively represent the right- and left-handed polarizations, and the polarization tensors $e_{ij}^A({\bf k})$ satisfy the relation $\epsilon_{ijk}k_i e_{kl}^A = ik \lambda_A e^A_{jl}$ with $\lambda_R=1$ and $\lambda_L=-1$. We then canonically quantize the fields $h^A(\eta,{\bf k})$ by expanding them as 
\begin{align}\label{quantization}
h^A(\eta,{\bf k}) = h^A_k(\eta) \hat{a}^A_{\bf k} +  h^{\ast A}_k(\eta) \hat{a}^{\dagger A}_{\bf -k}\,,
\end{align}
where $\hat{a}^A_{\bf k}$ and $\hat{a}^{\dagger A}_{\bf k}$ denote the annihilation and creation operators satisfying the commutation relation  $[\hat{a}^A_{\bf k},\hat{a}^{\dagger A^\prime}_{\bf k^\prime}] = \delta^{A A^\prime}\delta({\bf k - k^\prime})$. 
Substituting Eqs. \eqref{Fourier} and \eqref{quantization} into Eq. \eqref{EoM_h}, the mode function $h^A_k$ can be shown to obey
\begin{align}\label{EoM_h_k}
\ddot h^A_k + 3H \dot h^A_k + \frac{k}{a} \left(\frac{k}{a} + c\lambda_A \dot \theta  \right)h^A_k = 0\,.
\end{align}
The term in \eqref{EoM_h_k} proportional to $c$ induces the velocity birefringence phenomenon during the propagation of GWs. More interestingly, this term can trigger the tachyonic instability of one of the two polarization states, because the frequency $\omega^A_k \equiv \sqrt{k^2/a^2 + c\lambda_A \dot\theta k/a} $ could lead to  a state where $(\omega^{A}_k)^2 < 0$. Without loss of generality, we can take $c>0$ and in general consider $\dot \theta < 0$. In this case, the right-handed polarization state experiences a tachyonic instability for $k/a<|c\dot\theta|$, bringing on an exponential growth of the perturbations.
If applying this model to the inflation, we may anticipate the parity-violating term to trigger the interesting phenomenological effect on the inflationary GW background, which will be explored in the next section.
  
\section{Axion inflation and chiral gravitational waves}\label{sec3}
\begin{figure}
	\centering
	\includegraphics[width=0.8\textwidth ]{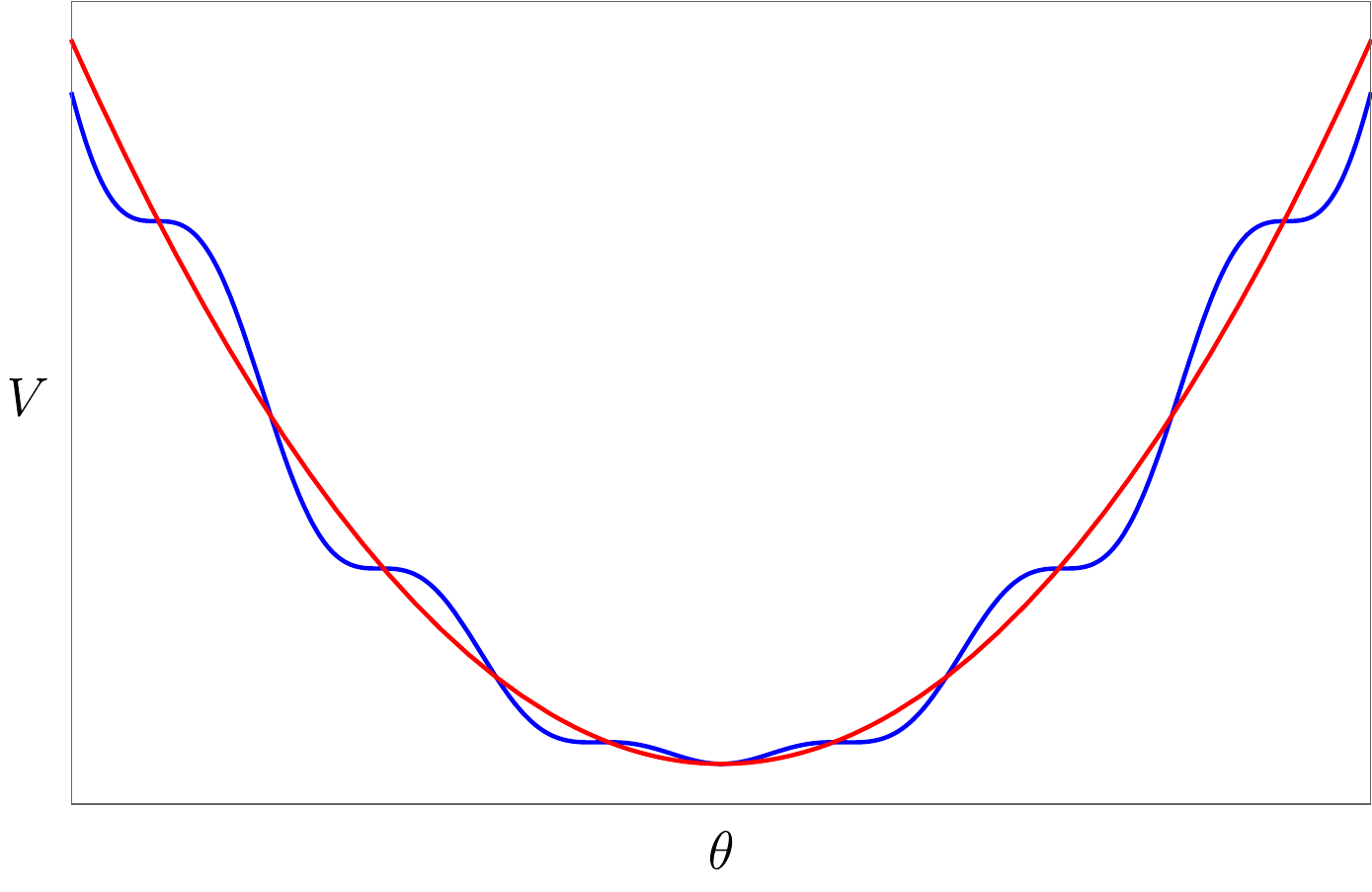}
	\caption{\label{fig1} Schematic diagram of the quadratic potential before (red line) and after (blue line) including sizable modulation.}
\end{figure}

Now we assume that the $\theta$ field plays the role of the inflaton to drive the cosmic inflation.
In the light of axion-like $\theta$ field, we consider a string-inspired model of axion inflation, where the potential has a monomial term plus sinusoidal modulation \cite{Kobayashi:2015aaa,CaboBizet:2016uzv}, 
\begin{align}\label{potential}
V(\theta) = \frac{1}{2}m^2\theta^2 + \Lambda^4 \frac{\theta}{f}\sin\left(\frac{\theta}{f}\right).
\end{align}
It is interesting to note that the sub-leading, but significant non-perturbative modulation effect with $\Lambda^4\lesssim m^2f^2$ can superimpose steep cliffs and gentle plateaus onto the quadratic potential, as seen in Fig. \ref{fig1}. These step-like structures of the potential could give rise to the non-trivial inflationary dynamics: the inflaton starts to slowly roll down a smooth plateau-like region, and then when it meets the sharp cliff connected by a flat plateau, the inflaton speeds up and quickly rolls through the cliff,  until it reaches the next plateau where Hubble friction rapidly slows down its motion \cite{Parameswaran:2016qqq}. Recent studies have shown that such a kind of the inflationary dynamics can lead to rich phenomenology. On the one hand, the flat plateau can result in so-called ultra-slow-roll inflation to realize the amplification of the primordial curvature perturbations at sub-CMB scales, which could generate a sizable amount of primordial black holes and an observable GW background \cite{Ozsoy:2018flq}. On the other hand, in the presence of coupling to gauge fields, the fast roll of the inflaton on the cliff-like region can trigger a localized production of the gauge field fluctuations that in turn sources the scalar and tensor perturbations \cite{Ozsoy:2020kat}. Next we study in detail the phenomenological consequences when the Nieh-Yan modified TEGR meets the axion inflation with the wiggly potential.

\begin{figure}
	\centering
	\includegraphics[width=0.85\textwidth ]{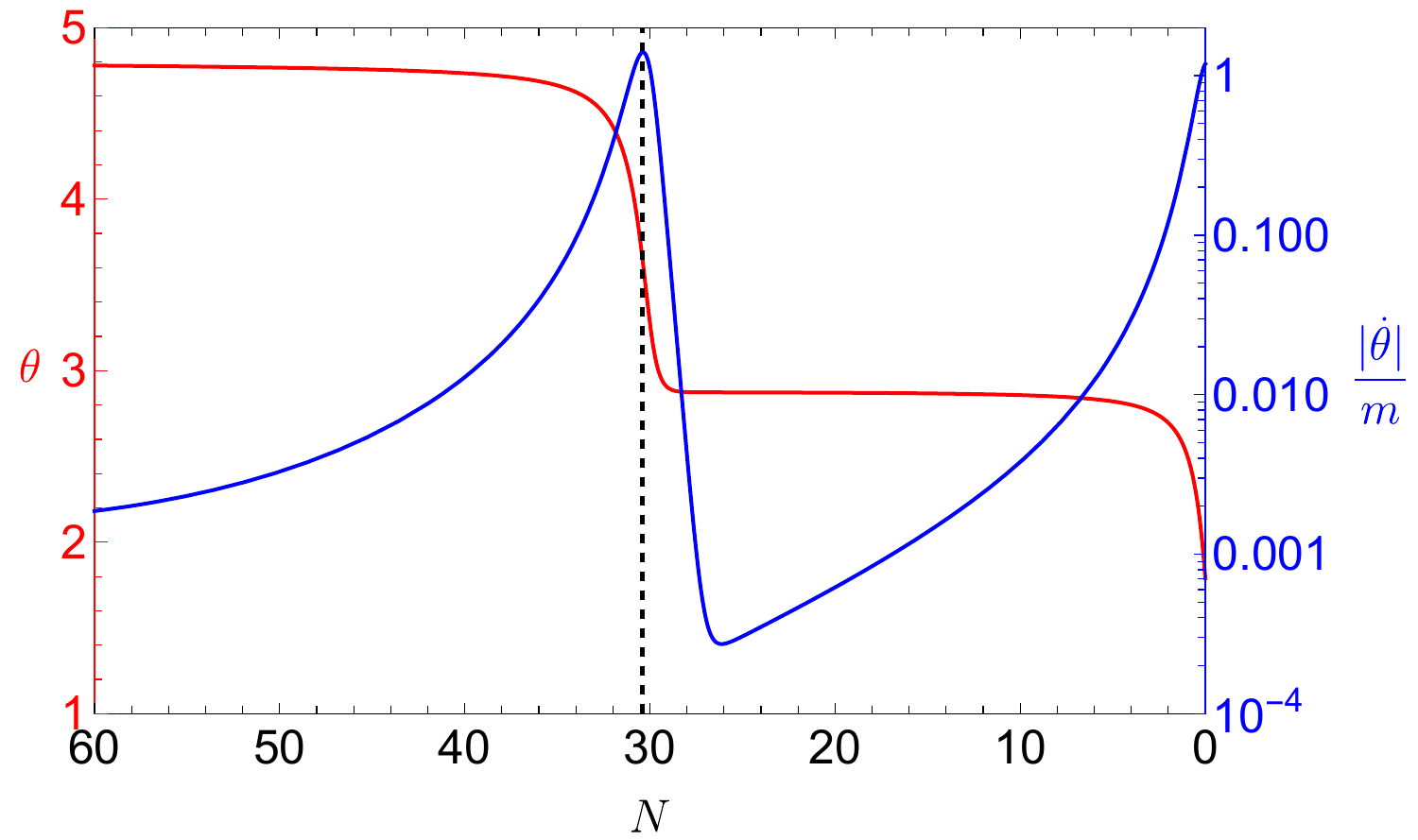}
	\caption{\label{fig2} The evolution of $\theta$ (red line) and $|\dot\theta|/m$ (blue line) as a function of \textit{e}-folding number $N$.}
\end{figure}

As an concrete example, we consider the following choice of parameters given in the axion potential \eqref{potential},
\begin{align}
\beta \equiv \frac{\Lambda^4}{ m^2f^2} = 0.996, \quad \alpha \equiv \frac{1}{f} = 3.295.
\end{align}
Then, by numerically solving the background equations given in Eq. \eqref{BG_Eq}, where the contribution from matter has been ignored, we display the evolution of $\theta$ and $|\dot\theta|/m$ as a function of the \textit{e}-folding number $N \equiv \ln(a_{\rm end}/a)$, with $a_{\rm end}$ being the scale factor at the end of inflation, in Fig. \ref{fig2}. From this figure, one can observe that the slow-roll evolution of the $\theta$ field in two smooth plateaus connected by the steep cliff takes up quiet a good deal of the whole $60$ \textit{e}-folds, and at around $N=30$, the $\theta$ field quickly descends over the cliff-like region, where the velocity $\dot\theta$ transiently peaks as $\theta$ accelerates down the cliff and then decelerates into the plateau. Before examining the effects of the presence of a large peak in the profile of $\dot\theta$ on the tensor perturbations, we must first determine the origin of the primordial scalar perturbations since $\zeta_\theta = 0$.
A feasible scenario is to consider that the matter described by $S_m$ is a curvaton field, which generates the primordial scalar perturbations instead of the inflaton. We carry out a comprehensive analysis of the curvaton scenario in Appendix \ref{appendix}. Next, we move on to the evolution of the tensor perturbations.

\begin{figure}
	\centering
	\includegraphics[width=0.9\textwidth ]{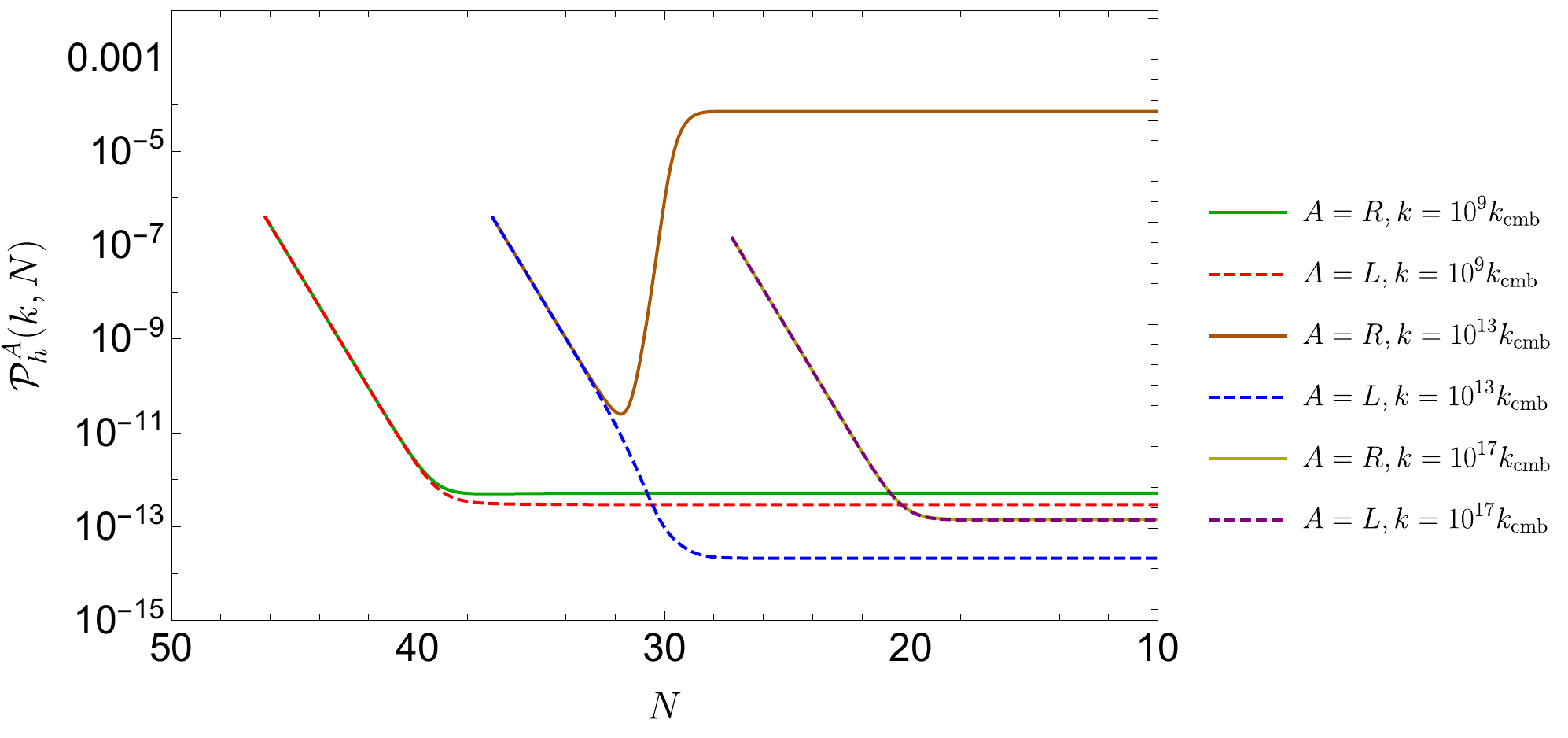}
	\caption{\label{fig3} The evolution of the power spectrum $\mathcal{P}^A_h\equiv k^3|h^A_k|^2/(2\pi^2)$ as a function of the \textit{e}-folding number $N$ for different $k$ modes: $k=10^9k_{\rm cmb}$, $k=10^{13}k_{\rm cmb}$, and $k=10^{17}k_{\rm cmb}$.}
\end{figure}

\begin{figure}
	\centering
	\includegraphics[width=0.48\textwidth ]{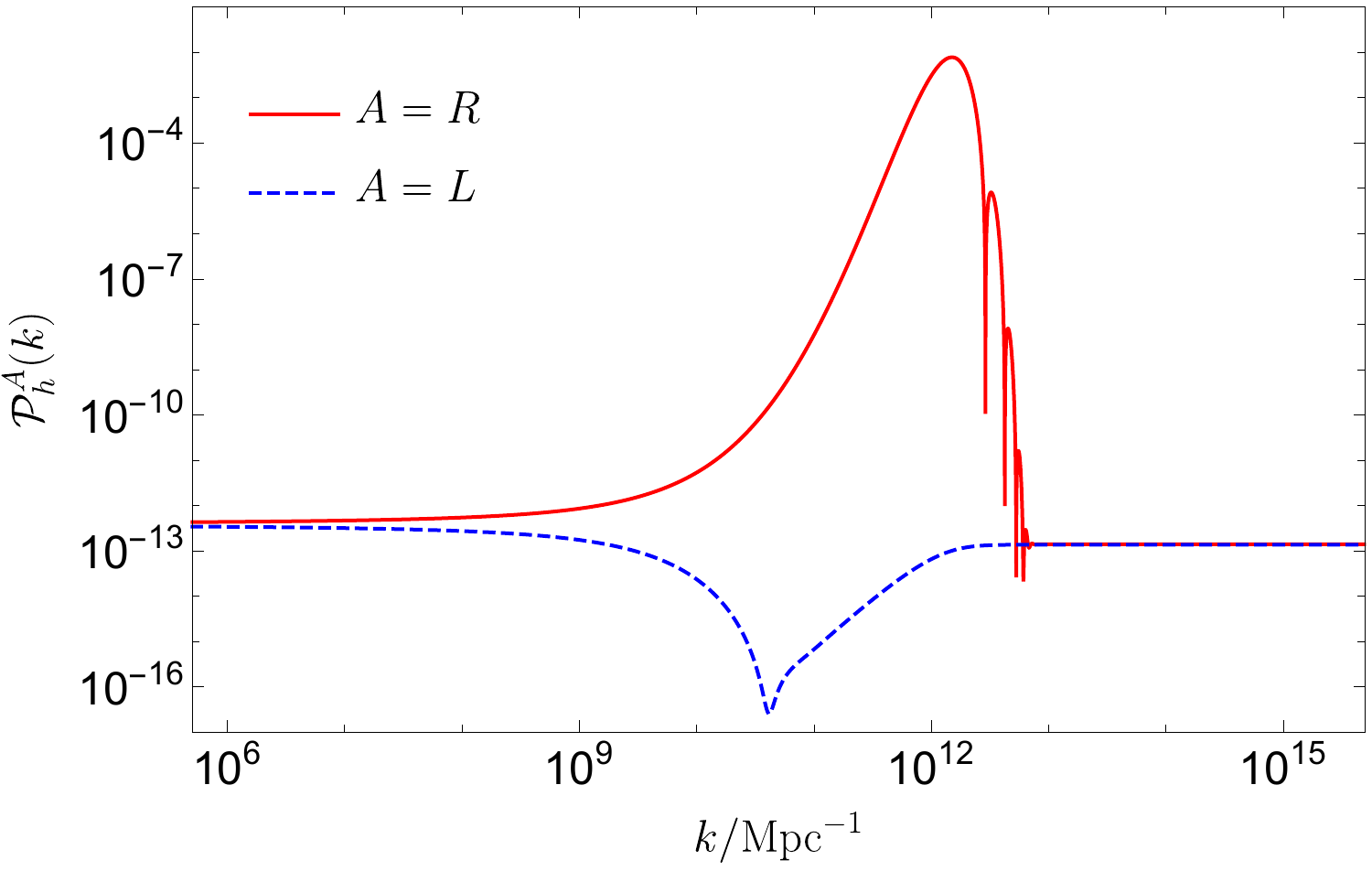}
	\includegraphics[width=0.48\textwidth ]{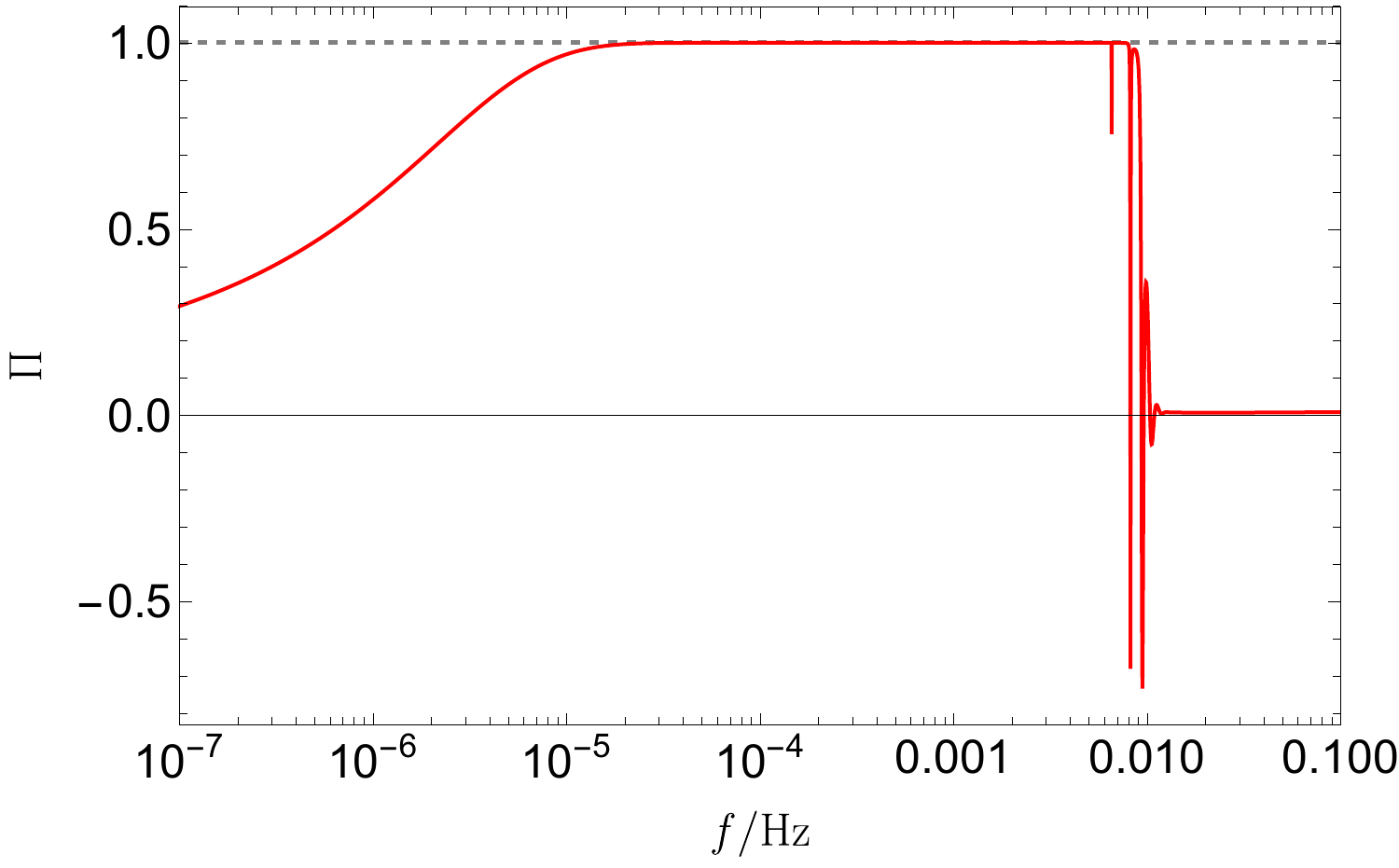}
	\caption{\label{fig4} Left: the power spectra for right- and left-handed polarizations. Right: the polarization degree $\Pi$ as a function of frequency $f$. }
\end{figure}

Combining with the background evolution shown in Fig. \ref{fig2}, we take $c=20$ and set the \textit{e}-folds $N$ at the time when the CMB scale $k_{\rm cmb}$ exits the horizon as $60$. In addition, we choose $m=10^{-6}$ as an example since the inflationary scale is lack of efficient constraint in the curvaton scenario (see Appendix \ref{appendix}).
Figure \ref{fig3} shows the evolution of the power spectrum $\mathcal{P}^A_h\equiv k^3|h^A_k|^2/(2\pi^2)$ as a function of the \textit{e}-folding number $N$ for different $k$ modes, in which the mode with $k=10^{13}k_{\rm cmb}$ corresponds to the frequency scale of LISA and Taiji. It is interesting to observe that the power spectrum only for the right-handed polarization state with $k=10^{13}k_{\rm cmb}$ experiences an exponential growth caused by the tachyonic instability.
Actually during inflation, the tachyonic instability has a significant influence on the right-handed polarization state only if it starts inside the horizon, i.e. $|c\dot\theta|\gtrsim \mathcal{O}(H)$ when $k/a <|c\dot\theta|$ is just beginning to be met. For the inflationary dynamics considered here, the case of $|c\dot\theta|\gtrsim \mathcal{O}(H)$ occurs exclusively during the transient fast-roll period of the $\theta$ field. On the one hand, although the condition of $k/a <|c\dot\theta|$ for the modes (e.g. $k=10^9k_{\rm cmb}$) that cross the horizon far before $N=30$ is satisfied during the fast-roll phase, the corresponding mode functions $h^R_k$ remain constant values since the frequencies $\omega^R_k$ are negligible compared with $H$. On the other hand, for the modes, e.g. $k=10^{17}k_{\rm cmb}$, which are still deep inside the horizon around $N=30$, the system has entered into the slow-roll phase when $k/a <|c\dot\theta|$, and thus the evolution of the corresponding mode functions doesn't feel the appearance of the parity-violating term. As a result, the tachyonic instability induced by the parity-violating term leads to an efficient enhancement of the amplitude of $h^R_k$ with certain modes, e.g. $k=10^{13}k_{\rm cmb}$, whose horizon-crossing time is near or within the fast-roll phase. This phenomenon finally yields a large localized bump in the power spectrum for the right-handed polarization state, as one can see from the left panel of Fig. \ref{fig4}. Moreover, the degree of the circular polarization, defined as  
\begin{align}
	\Pi = \frac{\mathcal{P}^R_h-\mathcal{P}^L_h}{\mathcal{P}^R_h+\mathcal{P}^L_h}\,,
\end{align}
is plotted in the right panel of Fig. \ref{fig4}. According to definition, $\Pi =1,-1$ correspond to fully right-handed and left-handed polarized GWs, respectively. It is clear that the resulting GW signal is almost fully right-handed polarized ($\Pi \simeq 1$) within the frequency range from $10^{-5}{\rm Hz}$ to $10^{-2}{\rm Hz}$.

\begin{figure}
	\centering
	\includegraphics[width=0.9\textwidth ]{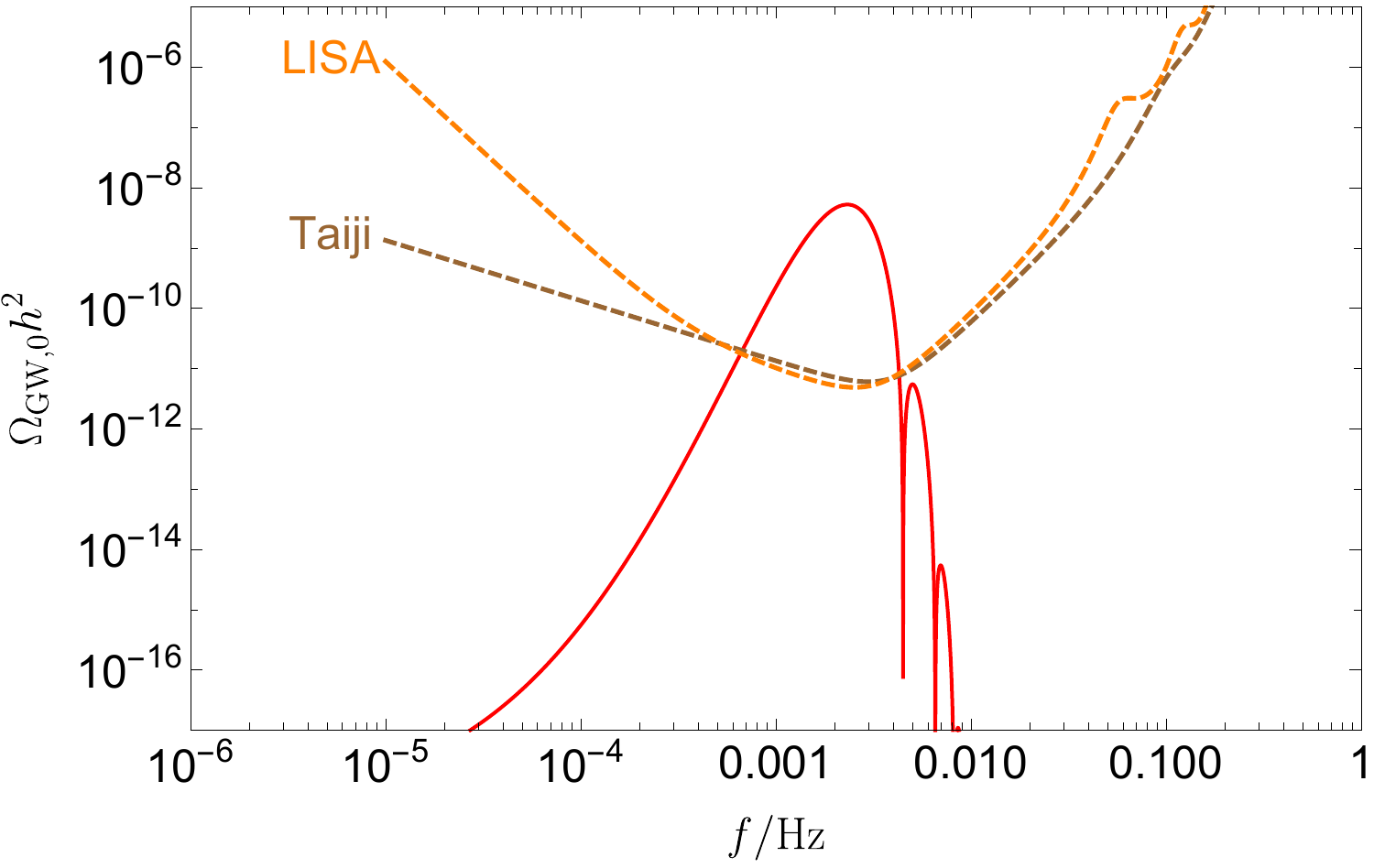}
	\caption{\label{fig5} The current energy spectrum $\Omega_{\rm GW,0}h^2$ of the predicted GW background. The orange and brown dashed lines represent the sensitivity curves of the space-based projects, LISA \cite{LISA:2017pwj} and Taiji \cite{Ruan:2018tsw}, respectively. }
\end{figure}

The current energy spectrum of GWs is related to the power spectrum generated during inflation through \cite{Inomata:2021zel}
\begin{align}
\Omega_{\rm GW,0}(k)h^2 = 6.8\times 10^{-7} \mathcal{P}_h(k)\,,
\end{align}
where $\mathcal{P}_h$ ($=\mathcal{P}^R_h+\mathcal{P}^L_h$) is the total power spectrum of the tensor perturbations, and we show the resulting $\Omega_{\rm GW,0}h^2$ in Fig. \ref{fig5}.
It is easy to see that the energy spectrum of the stochastic GW background exceeds the sensitivity curves of LISA and Taiji. 
In \cite{Seto:2020zxw,Orlando:2020oko} it was shown that a clear measurement of chirality can be claimed for a maximally chiral ($|\Pi|=1$) GW signal with $\Omega_{\rm GW}h^2 \sim 10^{-12}$ in the LISA-Taiji frequency range. Therefore, the chirality of the GW signal predicted in present paper can be detected through the LISA-Taiji network.
The measurable chirality of the predicted GW background will be a very distinctive property that allows to distinguish this model from other inflationary scenarios.

\section{conclusions}\label{sec4}
A simple and healthy parity-violating gravity model recently was constructed by incorporating the Nieh-Yan term coupled to an axion-like field into the TEGR action \cite{Li:2020xjt}.
By applying this model to the spatially flat FRW cosmology, it was shown that the presence of the Nieh-Yan term causes the gauge-invariant scalar perturbation $\zeta_\theta$ associated with the coupled scalar field $\theta$ to vanish identically, as well as the tensor perturbations to exhibit velocity birefringence. In this paper, we investigate the phenomenological implications of this model during inflation. We consider that the axion-like $\theta$ field has a potential with step-like features and acts as an inflaton to drive inflation. It should be emphasized that the $\theta$ field itself plays no part in generating the primordial scalar perturbations. However, we find that the curvaton scenario may be employed to account for the origin of the primordial scalar perturbations. For tensor perturbations, the parity-violating term inducing velocity birefringence can trigger the tachyonic instability of the right-handed (or left-handed) polarization state. Due to the fast roll of the $\theta$ field on the cliff-like region of its potential, the perturbations for the right-handed (or left-handed) polarization state, with the modes that exit the horizon around this phase, experience an exponential growth. As a consequence, the power spectrum for the tensor perturbations presents a large localized bump contributed by one polarization state. The resulting chiral GW background is detectable by LISA and Taiji, and its chirality can be determined by correlating two detectors. The measurement of this GW signal provides a powerful way to access information on inflation and test the model we study in this work.

\begin{acknowledgments}
We are grateful to Prof. Tao Zhu and Prof. Puxun Wu for useful discussions and comments. This work is supported in part by the National Key Research and Development Program of China Grant No. 2020YFC2201502, the National Natural Science Foundation of China Grants No. 11947302, No. 11991052, No. 11690022, No. 11821505 and No. 12047559, the China Postdoctoral Science Foundation Grant No. 2020M680689, the Key Research Program of the CAS Grant No. XDPB15, and the Key Research Program of Frontier Sciences of CAS.

\end{acknowledgments}

\appendix
\section{Curvaton scenario}\label{appendix}
In this appendix, we first briefly review the curvaton scenario, and then determine the conditions required to make the resulting scalar spectrum compatible with the current CMB observations.

For simplicity, we consider a minimally coupled canonical curvaton field, $\sigma$, specified by the action
\begin{align}
S_\sigma = \int dx^4 \sqrt{-g} \left[ \frac{1}{2}\nabla_\mu \sigma\nabla^\mu \sigma - U(\sigma) \right]\,.
\end{align}
Then, the equation of motion for the homogeneous component of the curvaton reads
\begin{align}
\ddot\sigma + 3H\dot\sigma + U_{\sigma} = 0\,,
\end{align}
and the mode function of the curvaton field perturbations satisfies 
\begin{align}\label{EoM_curvaton_pert}
	\delta\ddot\sigma_k + 3H\delta\dot\sigma_k  + \left[\frac{k^2}{a^2} + U_{\sigma\sigma}  - \frac{1}{a^3}\frac{d}{dt}\left( \frac{a^3}{H}\dot\sigma^2\right) \right]\delta\sigma_k = 0\,,
\end{align}
which is derived by using the linearized Einstein equations:
\begin{align}
6H^2\phi + 2H\frac{\nabla^2}{a^2}(aB) &=  ( \dot\theta^2 + \dot\sigma^2) \phi - \dot\sigma\delta\dot\sigma -  U_{\sigma}\delta\sigma\,,  \nonumber \\
2H \phi &=  \dot\sigma\delta\sigma\,, \nonumber  \\
2H \dot\phi + (4\dot H + 6 H^2) \phi  &= - (\dot\theta^2 + \dot\sigma^2 ) \phi + \dot\sigma\delta\dot\sigma -  U_{\sigma}\delta\sigma\,.
\end{align}
In the standard curvaton scenario, the curvaton is expected to be a light field ($|U_{\sigma\sigma}| \ll H^2$) and to have a subdominant energy density ($U \ll H^2$) during inflation, in addition, the initial curvaton value in general satisfies $H \ll \sigma \ll 1$. Using the slow-roll approximation $3H\dot\sigma \simeq - U_{\sigma} $, Eq. \eqref{EoM_curvaton_pert} can be simplified as  
\begin{align}\label{EoM_curvaton_pert_1}
\delta\ddot\sigma_k + 3H\delta\dot\sigma_k + \left( \frac{k^2}{a^2} + U_{\sigma\sigma}\right)\delta\sigma_k = 0\,.
\end{align}
It follows that on the superhorizon scales one can obtain a nearly scale-invariant power spectrum of the curvaton field perturbations,
\begin{align}
\mathcal{P}_{\delta\sigma}(k) = \left(\frac{H_\ast}{2\pi}\right)^2
\end{align}
with a tiny spectral tilt,
\begin{align}
n_{\sigma} \equiv \frac{d\ln \mathcal{P}_{\delta\sigma}}{d\ln k} = 2\frac{\dot H_\ast}{H_\ast^2} + \frac{2}{3}\frac{(U_{\sigma\sigma})_\ast}{H_\ast^2}\,,
\end{align}
where $\ast$ denotes the fact that quantity is evaluated at the epoch of horizon exit, $k=a_\ast H_\ast$.

We assume that at the end of inflation, the inflaton decays into radiation immediately, and meanwhile the curvaton keeps rolling slowly along its potential until $H \sim U_{\sigma\sigma}$, following which the curvaton starts oscillating around the minimum of its potential. Assuming that the potential $U(\sigma)$ is quadratic, $U=m_\sigma^2\sigma^2/2$, the curvaton behaves like the pressureless matter and then its time-averaged energy density is $\rho_\sigma = m_\sigma^2 \tilde{\sigma}^2/2$ with $\tilde{\sigma}$ being the amplitude of the oscillation. Furthermore, Eq. \eqref{EoM_curvaton_pert_1} holds at the oscillating phase due to $\sigma \ll 1$, such that the perturbations at superhorizon scales and the background share the same equation of motion. Therefore, the ratio $\delta\sigma/\sigma$ remains fixed on superhorizon scales, namely
\begin{align}
	\frac{\delta\sigma}{\sigma} = \left( \frac{\delta\sigma}{\sigma} \right)_\ast\,.
\end{align}
Before the curvaton decays, the content of Universe consists of the dominant radiation with the energy density $\rho_r$ ($\propto a^{-4}$) and the pressureless matter with $\rho_\sigma$ ($\propto a^{-3}$). Thus at such a point, the gauge-invariant scalar perturbation with the gauge $\psi=0$ can be written as 
\begin{align}\label{zeta}
\zeta = \frac{4\rho_r \zeta_r + 3\rho_\sigma\zeta_\sigma}{4\rho_r + 3\rho_\sigma}
\end{align}
with
\begin{align}
\zeta_r = - H\frac{\delta\rho_r}{\dot\rho_r}= \frac{1}{4}\frac{\delta\rho_r}{\rho_r}\,, \quad \zeta_\sigma = - H\frac{\delta\rho_\sigma}{\dot\rho_\sigma}= \frac{1}{3}\frac{\delta\rho_\sigma}{\rho_\sigma}\simeq\frac{2}{3}\frac{\delta\sigma}{\sigma} = \frac{2}{3}\left( \frac{\delta\sigma}{\sigma} \right)_\ast\,,
\end{align}
where $\zeta_r$ and $\zeta_\sigma$ are gauge-invariant scalar perturbations corresponding to the radiation and the curvaton, respectively. Since the radiation comes from the inflaton decay and $\zeta_\theta = 0$, we have $\zeta_r = 0$. Therefore, Eq. \eqref{zeta} becomes
\begin{align}
\zeta = \frac{ 3\rho_\sigma}{4\rho_r + 3\rho_\sigma} \zeta_\sigma = \frac{2}{3} r \left( \frac{\delta\sigma}{\sigma} \right)_\ast\,,
\end{align}
where $r\equiv 3\rho_\sigma/(4\rho_r + 3\rho_\sigma)$ denotes the radio of the curvaton to background energy density. The ratio $r$ changes with time until the curvaton decays into radiation, after which $\zeta$ becomes a constant. It follows that the final power spectrum of the scalar perturbations is 
\begin{align}\label{P_R}
\mathcal{P}_\zeta = \frac{r_{\rm dec}^2}{9\pi^2}\frac{H_\ast^2}{\sigma_\ast^2}\,,
\end{align}
where $r_{\rm dec}$ is $r$ at the decay epoch, and its spectral index is given by
\begin{align}\label{index}
n_s -1 = n_\sigma = 2\frac{\dot H_\ast}{H_\ast^2} + \frac{2}{3}\frac{(U_{\sigma\sigma})_\ast}{H_\ast^2}\,.
\end{align}

A realistic inflation must be consistent with present CMB observation results. The spectral index $n_s$ at $k_{\rm cmb}$ has been constrained to be $n_s = 0.9649\pm0.0042$ at $68\%$ confidence level \cite{Planck:2018jri}. In the specific case of the axion inflation studied in this paper, the Hubble slow-roll parameter $|\dot H_\ast/H_\ast^2|$ is negligible in comparison to $|n_s-1|$, so that we require that $(U_{\sigma\sigma})_\ast$ is negative. Based on this, we can consider that the curvaton field has the $\alpha$-attractor E-model potential \cite{Kallosh:2013hoa},
\begin{align}
U(\sigma) = \Lambda_\sigma^4 \left( 1- e^{-\sigma/M}\right)^2\,,
\end{align}
which has a negative second-order derivative when $\sigma > M\ln 2$ and behaves as $\sigma^2$ with the effective mass squared $m_\sigma^2=2\Lambda_\sigma^4/M^2$ in the regime of $\sigma \ll M$. As an example, we simply take $\sigma_\ast = M$ with $H \ll M \ll 1$ when $N=60$, and then we obtain $(U_{\sigma\sigma})_\ast \simeq -0.2 \Lambda_\sigma^4/M^2 $. Note that in this case, the potential $U$ is not quadratic at the onset of the oscillating period, but become nearly quadratic after a few Hubble times as the oscillation amplitude decreases. As a consequence, the ratio $\delta\sigma/\sigma$ still has some constant value,
\begin{align}
	\frac{\delta\sigma}{\sigma} = q \left( \frac{\delta\sigma}{\sigma} \right)_\ast\,,
\end{align}
where $q$ is a time-independent factor depending on the potential parameters and the initial field value. Then, the scalar spectrum \eqref{P_R} can be rewritten as 
\begin{align}
	\mathcal{P}_\zeta = \frac{r_{\rm dec}^2q^2}{9\pi^2}\frac{H_\ast^2}{\sigma_\ast^2}\,,
\end{align}
which has the same spectral index as that given in Eq. \eqref{index}. Finally, from the present observational constraints on $\mathcal{P_R}$ and $n_s$, we have
\begin{align}
 \frac{ H_\ast }{M} \simeq 4.3\times 10^{-4} (r_{\rm dec}q)^{-1}\,, \quad 0.49 \lesssim \frac{\Lambda_\sigma^2}{ M H_\ast } \lesssim 0.55\,,
\end{align}
which put a significantly weak limit on the inflationary energy scale.

\bibliographystyle{apsrev4-1}
\bibliography{references}

\end{document}